\documentclass[twocolumn,english,aps,pra,superscriptaddress,showpacs,tightenlines]{revtex4-1}
\usepackage{amsmath}
\usepackage{graphicx}
\usepackage{amssymb}

\begin{document}

\title{Weak-Value Amplification of light deflection by a dark atomic ensemble}

\author{Lan \surname{Zhou}}
\affiliation{Key Laboratory of Low-Dimensional Quantum Structures and Quantum
Control of Ministry of Education, and Department of Physics, Hunan
Normal University, Changsha 410081, China}
\author{Yusuf \surname{Turek}}% yusufu@itp.ac.cn
\affiliation{State Key Laboratory of Theoretical Physics, Institute of Theoretical
Physics, University of Chinese Academy of Science, Beijing 100190, China }
\author{C. P. \surname{Sun}}
\affiliation{Beijing Computational Science Research Center, Beijing 100084, China}
\author{Franco Nori}
\affiliation{Advanced Science Institute, RIKEN, Wako-shi 351-0198, Japan}
\affiliation{Physics Department, The University of Michigan, Ann Arbor, MI 48109-1040, USA.}

\begin{abstract}
We study the coherent propagation of light whose dynamics is governed by the effective
Schr\"{o}dinger equation derived in a magneto-optically-manipulated atomic ensemble with
a four-level tripod configuration for electromagnetically induced transparency (EIT).
The small transverse deflection of an optical beam, which is ultra-sensitive to the EIT
effect, could be drastically amplified via a weak measurement with an appropriate
preselection and postselection of the polarization state. The physical mechanism is
explained as the effect of wavepacket reshaping, which results in an enlarged group
velocity in the transverse direction.
\end{abstract}

\maketitle

\narrowtext

\section{Introduction}

The weak measurement proposed by Aharonov, Albert and Vaidamn (AAV)~\cite{WVPRL60,qparadox}
is usually referred to an amplification effect for weak signals rather than a conventional
quantum measurement that collapses a coherent superposition of quantum states~\cite{VonNm}.
In the original Gedanken experiment~\cite{WVPRL60}, a seemingly surprising effect was
proposed as it is possible to ``measure'' an appropriate preselected state of the spin-$1/2$
particle to obtain the average spin beyond the bound $\left[ -0.5,0.5\right] $ of the smallest
and largest eigenvalues. Such average spin (called weak value) does not exactly reflect the
reality of a single spin, and is only the ensemble average of the measured data over a postselected
subset. In this sense, the weak measurement implies some amplification of the resulting
signal in the apparatus carrying on the pre- and post- selections~\cite{WVpr12}. Several
groups have studied such weak value amplification (WVA) effect in various physical systems,
such as quantum optical~\cite{phoPRL91,phoPRL05,SagPRL09,GCGuoA06} and condensed matter
systems~\cite{WilRL08,RomPRL08}. Note that WVA has been utilized as a quantum-coherent manipulation
to engineer various interesting quantum effects, such as superluminal effects in birefringent
optic fiber~\cite{lumPRL04}, ultra-sensitive beam deflection~\cite{SagPRL09} and optical
spin-Hall effects~\cite{Bliokh06,OSHE08,OSHE08F,OSHE08n,Gorod12}.

In this paper, we study the weak-value problem of spin-$1/2$ with polarized light beams
propagating in a dispersive medium, a four-level atomic ensemble controlled by external
fields. The motion of the light-wave envelope in such a coherent medium is described by
an effective Schr\"{o}dinger equation resembling the precession of a spin-$1/2$ in an
inhomogeneous magnetic field~\cite{DLZhou07,ZhouA08,ZhangA09,GuoA08}. However, because
the effective magnetic field is very weak due to the inhomogeneity of the coupling
transitions, the deflection of the light beam is usually difficult to observe even for
EIT with two-photon resonance. For example, in the experiment by Karpa and Weitz~\cite{Karpa06},
the angle of deflection is only about $2\times10^{-5}$ rad when the light passes through
a gas cell $5$cm long.

Since our concern here is only the deflection of the optical beam rather than the
enhancement of the beam split, and the two beam's split may be deflected in the
same direction, we consider weak measurements to achieve this goal. According to
our previous series of studies~\cite{DLZhou07,ZhouA08,GuoA08,ZhangA09}, the
motion of the transverse wavepacket is described by a Schr\"{o}dinger-like
equation, similar to that for the spin-$1/2$ in a transversely-inhomogeneous
magnetic field in the Stern-Gerlach experiment~\cite{GuoA08}. Thus, the peak
of the optical beam will split into two, according to the initial polarization,
under the influence of the transverse magnetic field gradient. When making a
postselection on the final polarized state, the projection on this chosen
polarization state mixes the two local wave packets in the transverse split.
If such a weak measurement does not deform too much the shape of wavepacket, the
transverse distance of the reshaped wavepacket from the center of the beam may
be very large compared to that for any peaks in the optical Stern-Gerlach experiment.
We carry out the relevant calculations in detail by making use of the effective
field approach~\cite{Lukin,Lukin1}.

This paper is organized as follows. In Sec.~\ref{Sec:2}, we present a
theoretical model for a four-level atomic ensemble with tripod
configuration in the presence of nonuniform external fields, and derive the
system of equations for the dynamics of the signal field in the atomic linear
response with respect to the probe field. In Sec.~\ref{Sec:3}, we present
the optical Stern-Gerlach effect of the probe beam in momentum space, which is
induced by the interaction between the dispersive atomic medium and probe
field used for EIT. In Sec.~\ref{Sec:4}, we theoretically study the transverse
deflection of the probe field via weak measurements, which can amplify signals
by appropriate preselected and postselected states of the system. We conclude
our paper in the final section.

\section{\label{Sec:2}model setup}

\begin{figure}[tbp]
\includegraphics[bb=51 468 392 751, width=8 cm]{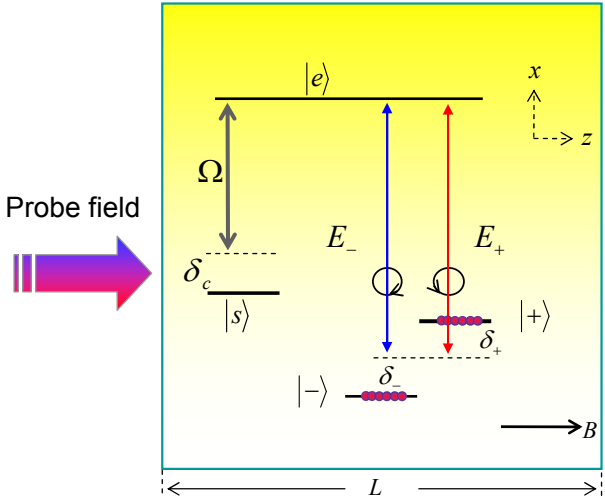}
\caption{(Color online) Schematic diagram of an atomic ensemble with a
four-level tripod configuration in a gas cell of length $L$, manipulated
by two optical and one magnetic fields. The magnetic field $B$ is
applied along the $z$-direction with a gradient along the $x$-direction.}
\label{fig2:1}
\end{figure}

Consider an ensemble of $N$ identical and noninteracting atoms confined in a
rectangular gas-cell, characterized by the ground-state Zeeman sublevels $%
\left\vert \pm \right\rangle $, one intermediate state $\left\vert s\right\rangle $,
and an excited state $\left\vert e\right\rangle $, as shown in Fig.~\ref{fig2:1}.
The levels are coupled by two optical fields: a control laser field with frequency
$\nu _{c}$ and wave-number $k_{c}$, and a probe laser field with frequency $\nu $
and wave-number $k$. The control field is assumed to be homogeneous and strong
enough for propagation effects to be neglected, which was tuned to the
$\left\vert s\right\rangle \rightarrow\left\vert e\right\rangle $ transition.
The coupling strength is characterized by the Rabi frequency $\Omega $. The
probe laser field is linearly-polarized and propagating along the $z$-axis. Its
linear polarization is a superposition of left- and right-handed circular
polarization, labelled by $\sigma _{\pm }$. We denote the $\sigma _{j}$-polarized
component as $\tilde{E}_{j}\left( \mathbf{r},t\right) ,j=\pm $, which drives
the transitions $\left\vert \pm \right\rangle \leftrightarrow \left\vert
e\right\rangle $, respectively.

The atomic gas cell can be divided into many smaller cells. We assume that
each smaller cell contains a large number of atoms and the inhomogeneous
external field is sufficiently homogeneous for each smaller cell~\cite%
{DLZhou07}. In this case, the atomic medium can be treated in a continuous
way by the following approach. First, describe the medium excitation by
introducing the collective atomic operators $\tilde{\Xi}_{\mu \nu }\left(
\mathbf{r}\right) =(1/N_{\mathbf{r}})\sum_{j=1}^{N_{\mathbf{r}}}\tilde{\Xi}%
_{\mu \nu }^{j}$, averaged over a small but macroscopic volume
containing many atoms $N_{\mathbf{r}}=(N/V)dV\gg 1$ around position $\mathbf{%
r}$. Here, $\tilde{\Xi}_{\mu \nu }^{j}\left( \mathbf{r}\right) =\left\vert
\mu \right\rangle _{j}\langle \nu |$. Afterwards replace the sum over the total
number $N$ of atoms by $\frac{N}{V}\int d^{3}r$, where $V$ is the volume
of the medium~\cite{Lukin,Lukin1}. Neglecting the kinetic energy of the atoms,
the Hamiltonian of the atomic part is given by
\begin{equation}
H^{(A)}=\frac{N}{V}\sum_{i}\int d^{3}r\left( \omega _{i}+\mu _{i}B\right)
\tilde{\Xi}_{ii}  \label{SecII-01}
\end{equation}%
where $\omega _{i}$ $(i=\pm ,s,e)$ are the bare atomic energies, and $\omega
_{\pm }=\omega _{0}$ (which corresponds to degenerate levels when $B=0$). The
magnetic field $B$ along the $z$-axis shifts the energy levels
by the amount $\mu _{i}B\left( \mathbf{r}\right) $, where the
magnetic moments, $\mu _{i}=m_{F}^{i}g_{F}^{i}\mu _{B}$, are defined by the
Bohr magneton $\mu _{B}$, the gyromagnetic factor $g_{F}^{i}$, and the
magnetic quantum number $m_{F}^{i}$ of the corresponding state $\left\vert
i\right\rangle $. Under the electric-dipole approximation and the
rotating-wave approximation, the light-matter interaction Hamiltonian becomes~%
\cite{Lukin,Lukin1}
\begin{eqnarray}
H^{(I)} &=&\frac{N}{V}\int \Omega\; e^{i\left( \mathbf{k}_{c}\cdot \mathbf{r}%
-\nu _{c}t\right) }\;\tilde{\Xi}_{es}\;d^{3}r+h.c.  \label{SecII-02} \\
&&+\frac{N}{V}\sum_{j=\pm }d_{ej}\int \tilde{E}_{j}^{\left( +\right) }\;\tilde{%
\Xi}_{ej}\;d^{3}r+h.c.,  \notag
\end{eqnarray}%
with $d_{e-}(d_{e+})$ denoting the matrix element of the dipole momentum operator
projected on the direction of the electric field.

The slow-varying variables $E_{j}(\mathbf{r},t)$ for the probe field and the collective
atomic transition operators $\Theta _{\alpha \beta }$ can be defined as
\begin{subequations}
\label{SecII-03}
\begin{eqnarray}
\tilde{E}_{j}^{+}(\mathbf{r},t) &=&\sqrt{\frac{\nu }{2\varepsilon _{0}V}}%
E_{j}(\mathbf{r},t)e^{i(kz-\nu t)},\;(j=\pm )\text{,} \\
\tilde{\Xi}_{ej} &=&\Theta _{ej}\exp (-ikz)\text{,} \\
\tilde{\Xi}_{es} &=&\Theta _{es}\exp (-i\mathbf{k}_{c}\cdot \mathbf{r})\text{%
.}
\end{eqnarray}%
In a rotating reference frame, the dynamics of this system is described by
\end{subequations}
\begin{eqnarray}
H_{{\rm rot}} &=&\frac{N}{V}\int d^{3}r\left( \delta _{-}\Theta _{--}+\delta
_{+}\Theta _{++}+\delta _{c}\Theta _{ss}\right)   \label{SecII-04} \\
&&+\frac{N}{V}\int d^{3}r\left( \Omega \Theta _{es}+\sum_{j=\pm
}g_{j}E_{j}\Theta _{ej}\right) +h.c.  \notag
\end{eqnarray}%
with $\delta _{i}$ $(i=\pm ,c)$ the detunings of the probe and control fields
from the corresponding atomic transitions given as
\begin{align*}
\delta _{c}& =\omega _{s}-\omega _{e}+\nu _{c}+\left( \mu _{s}-\mu
_{e}\right) B, \\
\delta _{\pm }& =\omega _{\pm }-\omega _{e}+\nu +\left( \mu _{\pm }-\mu
_{e}\right) B,
\end{align*}%
and the coupling strength as
\begin{equation}
g_{j}=d_{ej}\sqrt{\frac{\nu }{2\varepsilon _{0}V}}.  \label{SecII-05}
\end{equation}%
The Hamiltonian in Eq.~(\ref{SecII-04}) generates the Heisenberg equations of the
slowly-varying variables $\Theta _{\mu \nu }$. The Langevin equations are
obtained to describe the dynamics of the medium by introducing the coherence
relaxation rate $\gamma $ between the ground state $|\pm \rangle $ and
the intermediate state $\left\vert s\right\rangle $, as well as the decay rate $%
\Gamma $ of the excited state. The low intensity approximation~\cite%
{Lukin,Lukin1,SunPRL01}, on one hand, allows us to neglect Langevin noise
operators since the number of photons contained in the probe laser beams is
much smaller than the number of atoms in the sample, so the operators
become c-numbers. On the other hand, it allows us to regard the interaction
between the matter and the probe field as a weak disturbance, since the
intensity of the probe laser beams is much weaker than that of the control
laser field. The perturbation approach~\cite{Lukin,Lukin1,Petrosyan} can be
applied in terms of a power series in $gE_{j}$:
\begin{equation}
\Theta _{\alpha \beta }=\Theta _{\alpha \beta }^{(0)}+\varepsilon \Theta
_{\alpha \beta }^{(1)}+\cdots   \label{SecII-06}
\end{equation}%
where $\varepsilon$ is a parameter that ranges continuously between zero
and one. When $\varepsilon=0$, the probe field is absent. For all atoms
initially in level $|\pm \rangle $ without polarization (i.e., atom $i$ in a
mixed state $\rho _{i}=\Sigma _{j=\pm }|j\rangle \langle j|/2$), we obtain
\begin{equation}
\Theta _{--}^{(0)}=\Theta _{++}^{(0)}=\frac{1}{2}  \label{SecII-07}
\end{equation}%
while all others terms vanish. Here, we retain only terms up to the first-order
in $\varepsilon $, since the linear optical response theory can sufficiently
reflect the main physical features. The dispersion and absorption are
determined by $\Theta _{je}^{(1)}$, which is obtained as
\begin{equation}
\Theta _{je}^{\left[ 1\right] }=-\frac{\Delta _{j}\,g_{j}\,E_{j}}{2\Omega\,\Omega
^{\ast }},\,\,(j=\pm )  \label{SecII-8}
\end{equation}%
in the steady-state solutions~\cite{GuoA08,ZhangA09,Petrosyan}, where pure
dephasing processes and decay among the lower states are neglected ($\gamma
=0$) to highlight the main physics, and a sufficiently strong
driving field is assumed to satisfy $\left\vert \Omega \right\vert ^{2}\gg
\Gamma \gamma ,\Delta _{j}^{2}$. We note the assumption on the strong
driving field implies that\emph{\ }$\left\vert \Omega \right\vert \gg
\left\vert \left( \mu _{+}-\mu _{-}\right) B\right\vert $. The Raman
detuning is defined as
\begin{equation}
\Delta _{j}=\delta _{j}-\delta _{c},  \label{SecII-9}
\end{equation}%
which leads to a spatially-varying refractive index profile in the gas
cell~\cite{DLZhou07} due to the small transverse magnetic field gradient.

Using the slowly-varying-envelope approximation, the paraxial wave equation
in the linear optical response theory~\cite{Lukin,Lukin1,Petrosyan}
\begin{equation}
\left( i\partial _{t}+ic\partial _{z}\right) E_{j}=-Ng_{j}^{\ast }\Theta
_{je}^{(1)},  \label{SecII-10}
\end{equation}%
becomes an effective Schr\"{o}dinger equation
\begin{equation}
i\partial _{t}E_{j}=H_{j}E_{j},  \label{SecII-11}
\end{equation}%
by substituting Eq.~(\ref{SecII-8}) into (\ref{SecII-10}), where the
effective Hamiltonian is
\begin{equation}
H_{j}=cp_{z}+\frac{N\left\vert g_{j}\right\vert ^{2}}{2\left\vert \Omega
\right\vert ^{2}}\Delta _{j}.  \label{SecII-12}
\end{equation}%
Here $c$ is the velocity of light in vacuum and $p_{z}=-i\partial _{z}$.
Notice that the $\sigma _{\pm }$-polarization accompany the component $%
E_{\pm }$. Writing the $\sigma _{\pm }$-polarization states as column
vectors resembling the spin-$1/2$
\begin{subequations}
\label{SecII-13}
\begin{eqnarray}
\left\vert \sigma _{+}\right\rangle &=&\left[
\begin{array}{cc}
1 & 0%
\end{array}%
\right] ^{T}, \\
\left\vert \sigma _{-}\right\rangle &=&\left[
\begin{array}{cc}
0 & 1%
\end{array}%
\right] ^{T},
\end{eqnarray}%
where the superscript $T$ means transpose, we can group the two components $%
E_{\pm }$ into a column vector defined as the \textquotedblleft spinor
state\textquotedblright\ $\Phi =\left[ E_{+},E_{-}\right]^T $. The dynamical
equation of the probe laser field reads
\end{subequations}
\begin{equation}
i\partial _{t}\Phi =\left[
\begin{array}{cc}
H_{+} & 0 \\
0 & H_{-}%
\end{array}%
\right] \Phi =H_{{\rm eff}}\Phi ,  \label{SecII-14}
\end{equation}%
which allows us to write the state of the probe laser field at any arbitrary time
as
\begin{equation}
\left\vert \Phi \left( t\right) \right\rangle =\sum_{j=\pm }c_{j}\left\vert
E_{j}\left( t\right) \right\rangle \left\vert \sigma _{j}\right\rangle .
\label{SecII-15}
\end{equation}%
Here, the states $\left\vert E_{j}\right\rangle $ describe the state of the
spatial degrees of freedom, with $E_{j}\left( \mathbf{r},t\right) $ referred
to as the corresponding spatial representation.

Hereafter, to investigate the beam deflection amplification of light beam,
propagating in the dispersive atomic ensemble, we use a signal
enhancement technique known from weak measurements~\cite{Aharonov90}. Along
with the standard weak measurement terminology, the transverse position
degree of freedom of the probe beam is referred to as the meter and its
intrinsic polarization degree of freedom is referred to as the measured
system.

\section{\label{Sec:3}optical Stern-Gerlach effect}%in momentum space}

We now investigate the evolution of the probe wave packet. The polarization
vector of the probe field lies in the plane perpendicular to its travelling
direction (i.e., the $z$-direction). It is initially prepared in a superposition
state
\begin{equation}
\left\vert i\right\rangle =\cos\frac{\alpha }{2}\left\vert H\right\rangle +\sin%
\frac{\alpha }{2}\left\vert V\right\rangle ,  \label{Sec3-01}
\end{equation}%
of the horizontal polarization, $|H\rangle =\left( \left\vert \sigma
_{+}\right\rangle +\left\vert \sigma _{-}\right\rangle \right) /\sqrt{2}$,
and vertical polarization, $|V\rangle =-i\left( \left\vert \sigma
_{+}\right\rangle -\left\vert \sigma _{-}\right\rangle \right) /\sqrt{2}$,
where $\alpha $ is the polarization angle of the probe light beam. For weak
measurement, this angle is very small, which means that the probe field is
initially almost in the horizontal polarization state. The role of $\alpha $
in the transverse beam deflection amplification via weak measurements
will be further discussed in the next section.

The two components of the probe field travel collinearly before reaching
the medium, which implies that initially $\left\vert E_{j}\left( t_{0}\right)
\right\rangle =\left\vert E_{0}\left( 0\right) \right\rangle $, where
the initial time $t_{0}=0$. After the probe field enters the
medium, the atomic ensemble induces the time evolution operator $%
U(t)=e^{-iH_{{\rm eff}}t}$ on the meter state according to their polarized state.
Then the state at an arbitrary time becomes an entangled state
\begin{equation}
\left\vert \Phi \left( t\right) \right\rangle =\frac{1}{\sqrt{2}}\left(
e^{-i\alpha /2}|\sigma _{+}\rangle |E_{+}(t)\rangle +e^{i\alpha /2}|\sigma
_{-}\rangle |E_{-}(t)\rangle \right) ,  \label{Sec3-02}
\end{equation}%
where the meter state is described by
\begin{equation}
\left\vert E_{j}(t)\right\rangle =e^{-iH_{j}t}\left\vert
E_{0}(0)\right\rangle .  \label{Sec3-03}
\end{equation}%
We will show that $E_{\pm }\left( \mathbf{r},t\right) $ in Eq.~$\left( \ref%
{Sec3-02}\right) $ implies a wavepacket split in momentum space
according to polarizations. % Thus we refer this phenomenon as the optical Stern-Gerlach effect in momentum space.
After a measurement on the postselected state $|V\rangle $, the meter is reshaped as
\begin{equation}
\left\vert \Phi _{f}^{m}\left( t\right) \right\rangle =\frac{i}{2}\left( e^{-i%
\frac{\alpha }{2}}\left\vert E_{+}(t)\right\rangle -e^{i\frac{\alpha }{2}%
}\left\vert E_{-}(t)\right\rangle \right) .  \label{Sec3-04}
\end{equation}%
Obviously, the superposition of the two wavepackets $E_{\pm }\left( \mathbf{r%
},t\right) $ can produce an interference pattern in the coordinate space.
\begin{figure}[tbp]
\includegraphics[width=8cm]{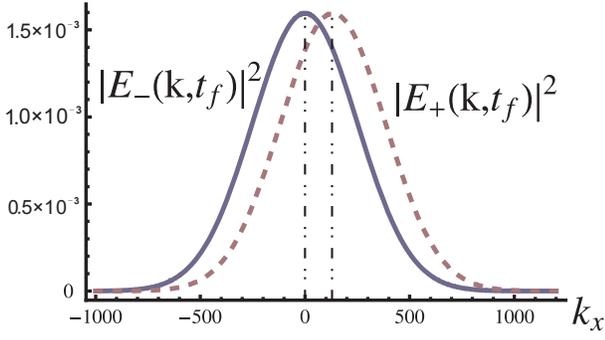}
\caption{(Color online) Optical Stern-Gerlach effect in momentum space with
the transverse distribution $\left\vert E_{j}\left( k_{x},t_{f}\right)
\right\vert ^{2}$ as a function of wave number $k_{x}=p_{x}/\hbar$, right
after the light leaves the EIT medium. Here, $k_{x}$ is in units of reciprocal
meters.}
\label{fig3:1}
\end{figure}

Now we assume that the magnetic field $B$ applied in the $z$-direction has a
linear gradient along the $x$-direction with the expression
\begin{equation}
B=B_{0}+B_{1}x.  \label{Sec3-05}
\end{equation}%
Here, our treatment is confined to only one transverse dimension, say
the $x$-direction. The effective Hamiltonian in Eq.(\ref{SecII-12}) reads
\begin{equation}
H_{j}=cp_{z}+b_{0j}+b_{1j}x,  \label{Sec3-06}
\end{equation}%
where the parameters
\begin{subequations}
\label{Sec3-07}
\begin{eqnarray}
b_{0j} &=&\frac{N\left\vert g_{j}\right\vert ^{2}}{2\left\vert \Omega
\right\vert ^{2}}\left[ \omega _{j}+\nu -\omega _{s}-\nu _{c}+\left( \mu
_{j}-\mu _{s}\right) B_{0}\right] , \\
b_{1j} &=&\frac{N\left\vert g_{j}\right\vert ^{2}}{2\left\vert \Omega
\right\vert ^{2}}\left( \mu _{j}-\mu _{s}\right) B_{1},
\end{eqnarray}%
can be adjusted by the control laser field and the magnetic field as well as
the energy levels. For an initial state with a two-dimensional Gaussian
amplitude profile
\end{subequations}
\begin{equation}
E_{j}\left( r,0\right) =\frac{1}{\sqrt{2\pi a^{2}}}\exp\left(-\frac{z^2+x^2}{4a^2}\right),  \label{Sec3-08}
\end{equation}%
the EIT medium introduces different phase shifts on the meter wavepacket
according to the right-handed or left-handed circularly-polarized state, and a
displacement $ct$ along the $z$-direction%
\begin{equation}
E_{j}\left( r,t\right) =\frac{e^{-itb_{0j}}}{\sqrt{2\pi a^{2}}}\exp\left[-\frac{(z-ct)^2+x^2}{4a^2}-itb_{1j}x\right].
\label{Sec3-09}
\end{equation}%
Since $\left[ H_{j},x\right] =0$, the center of $E_{j}\left( \mathbf{r}%
,t\right) $ does not change with time, $\left\langle x_{j}\right\rangle =0$,
which implies no spatial split of the meter wavepacket. A Fourier
transformation on Eq.(~\ref{Sec3-09})%
\begin{eqnarray}
\label{Sec3-10}
E_{j}\left( k,t\right) =&& 2\sqrt{2a^{2}\pi }\exp(-ib_{0j}t-a^{2}k_{z}^{2}-ick_{z}t) \\
&& \times\exp[-a^{2}\left(k_{x}+b_{1j}t\right) ^{2}] \notag
\end{eqnarray}%
shows that each meter's wavepacket keeps the longitudinal momentum unchanged
and acquires a transverse momentum with magnitude $b_{1j}t$, which is a
linear function of time. Therefore, two wavepackets of the probe field have
the same centroid, but achieve different momenta in the $x$-direction inside
the EIT medium. To illustrate the split of the probe beam in momentum space,
we assume that the probe beam with an initial width $a=2$ mm is tuned to the rubidium
($^{87}$Rb) D1-line $5^{^{2}}S_{_{1/2}}\leftrightarrow 5^{^{2}}P_{_{1/2}}$, that the
ground states $\left\vert \pm \right\rangle $ correspond to the magnetic
sublevels (with $m_{F}=1$ and $-1$) of the $F=1$ hyperfine ground state, and
that $\left\vert s\right\rangle $ represents the hyperfine ground state
$\left\vert F=2,m_{F}=1\right\rangle $. In this case, the phase shift on the
$\sigma _{-}$-component vanishes due to $\mu _{s}=\mu _{-}=4.64\times 10^{-24}\,
{\rm JT}^{-1}$. Hence, there is no shift of the momentum on the $\sigma_{-}$-component.
The magnetic field gradient $B_{1}=910\,\mu {\rm Gmm}^{-1}$
and the magnetic moments $\mu _{+}=-\mu _{s}$ subject the wavepacket of the $%
\sigma _{+}$-component to a linear potential in the transverse direction. In
Fig.~\ref{fig3:1}, we show such an optical Stern-Gerlach effect in momentum
space by plotting the transverse distribution $\left\vert E_{j}\left(k_{x},
t_{f}\right) \right\vert ^{2}$ as a function of the wavenumber $k_{x}=p_{x}/\hbar$
using the typical group velocities of a few thousand metres per second~\cite{KarpaPRL}
at a fixed time $t_{f},$ where $t_{f}=L/c$ denotes the interaction time with $L=50$
mm as the cell length. Obviously, the transverse displacement is smaller than
the uncertainty of the width of two wavepackets of the meter.

However, postselecting the system on a desired polarized state $\left\vert
V\right\rangle $ leads to a coherent superposition of two transverse
wavepackets. The interference of the two meter's wavepackets displace the
centroid of the wave packet in coordinate space by%
\begin{equation}
\left\langle x\right\rangle =\frac{\sin \left( b_{0}t+\alpha \right)
a^{2}b_{1}tf_{t}}{1-f_{t}\cos \left( b_{0}t+\alpha \right) },
\label{Sec3-11}
\end{equation}%
where we have introduced $b_{0}=b_{0+}-b_{0-}$, $b_{1}=b_{1+}-b_{1-}$ and $%
f_{t}\equiv \exp \left( -a^{2}b_{1}^{2}t^{2}/2\right) $. We also note that
when the optical beam is deflected by the EIT medium, the wave packet will
spread in the transverse direction. This spread will blur the observation of
deflection. To examine this, we calculate the transverse fluctuation%
\begin{equation*}
\left\langle \Delta x^{2}\right\rangle =a^{2}\frac{1+\left(
a^{2}b_{1}^{2}t^{2}-2\right) f_{t}\cos \alpha +\left( \cos ^{2}\alpha
-a^{2}b_{1}^{2}t^{2}\right) f_{t}^{2}}{\left( 1-f_{t}\cos \alpha \right) ^{2}%
}
\end{equation*}%
with $b_{0}=0$. Usually, only if $\langle\triangle x^{2}\rangle<\langle
x\rangle^{2}$, we can clearly observe such deflection. Otherwise, i.e., $%
\langle\triangle x^{2}\rangle>\langle x\rangle^{2}$, the deflection will be
blured by the transverse spreading of wavepackets. In this case, a weak-measurement
technique is necessary to observe the extremely small deflection of the
light beam.

\section{\label{Sec:4}weak value and deflection of the optical beam}

In the previous section, we have shown that the projection measurement on the
initial polarization described by the state $\left\vert i\right\rangle $ in
Eq.~(\ref{Sec3-01}), could induce the transverse displacement of the optical
beam after light passes through an EIT medium. Now, we will consider the
maximization of this transverse displacement, with the weak measurement
proposed by Aharanov et al.~\cite{WVPRL60}.

A weak measurement describes a situation where a system is so weakly coupled
to a measuring device that the uncertainty in the measurement is larger than
all the separations among the eigenvalues of the observable. Therefore, no
information is given since the eigenvalues are not fully resolved. Three
steps are necessary for weak measurements: 1) quantum state preparation
(preselection); 2) a weak perturbation; 3) postselection on a final quantum
state. The three essential ingredients of weak measurements were theoretically
performed in the previous section.

First, we have initially prepared the state $\left\vert i\right\rangle $ of
the system and a Guassian wave packet of the meter before the light is incident
on the medium. Second, the polarization-dependent effective potential in Eq.~(\ref{SecII-14})
changes the polarized state as
\begin{eqnarray}
 \label{Sec4-01}
\left\vert \phi \right\rangle =\frac{e^{-i\bar{\theta}_{x}t}}{\sqrt{2}}%
\left[ \exp\left(i\frac{\alpha +b_{0}t}{2}\right)\exp\left(i\frac{b_{1}xt}{2}\right)\left\vert
\sigma_-\right\rangle\right. \\
 +\left.\exp\left(-i\frac{\alpha +b_{0}t}{2}\right)
\exp\left(-i\frac{b_{1}xt}{2}\right)\left\vert \sigma_+\right\rangle \right] \notag
\end{eqnarray}%
with $\bar{\theta}_{x}=\left( \bar{b}_{0}+\bar{b}_{1}x\right) $, where $\bar{%
b}_{j}\equiv \left( b_{j+}+b_{j-}\right) /2$. The state $\left\vert \phi \right\rangle$
indicates that there is an interaction Hamiltonian between the system and the meter,
$H_{{\rm int}}=\xi\sigma_{z}$ with $\xi=b_{1}x/2$, and a free evolution $%
\exp\left(-ib_{0}t\sigma_{z}/2\right)$ on the spin. Here, $\left\vert
\sigma_{\pm} \right\rangle$ are the eigenstates of the Pauli operator $%
\sigma _{z}$ with eigenvalues $\pm 1$. A weak perturbation is guaranteed when
the transverse displacement in momentum space is smaller than the width
of the transverse distribution. In addition, the coupling strength $b_1$
could be adjusted by the control laser field and the magnetic field gradient.

Afterwards, the polarization state $|V\rangle $ is postselected. The
information on the observable of the system is read out from the transverse
spatial distribution which serves as the meter. However, the mean position
in Eq.~(\ref{Sec3-11}) is not the weak value. Weak measurements can provide
the weak values defined by $\left\langle \sigma_{z}\right\rangle _{w}=\left\langle
\psi _{{\rm fi}}\right\vert \hat{\sigma}_{z}\left\vert \psi _{{\rm in}}\right\rangle
/\left\langle \psi _{{\rm fi}}\right.\left\vert \psi _{{\rm in}}\right\rangle $,
where $\left\vert \psi_{{\rm in}}\right\rangle $ and $\left\vert \psi _{{\rm fi}}\right\rangle $
are the preselected and postselected states of the system, respectively. Here,
they are given by
\begin{subequations}
\label{Sec4-02}
\begin{eqnarray}
\left\vert \psi _{{\rm in}}\right\rangle \equiv &&\frac{e^{-i\bar{\theta}_{x}t}}{\sqrt{2}}
\left[ \exp\left(i\frac{\alpha +b_{0}t}{2}\right)\left\vert \sigma_-\right\rangle \right. \notag \\
&& \left.+\exp\left(-i\frac{\alpha +b_{0}t}{2}\right)\left\vert \sigma_+\right\rangle \right] , \\
\left\vert \psi _{{\rm fi}}\right\rangle &=&\left\vert V\right\rangle.
\end{eqnarray}
\end{subequations}
In our case, the observable is $\sigma _{z}$. Taking the free evolution of
the spin into account, we obtain the weak value
\begin{equation}
\left\langle \sigma _{z}\right\rangle _{w}=\frac{\left\langle \psi
_{{\rm fi}}\right\vert \sigma _{z}\left\vert \psi _{{\rm in}}\right\rangle }{%
\left\langle \psi _{{\rm fi}}\right. \left\vert \psi _{{\rm in}}\right\rangle }=i\cot
\left(\frac{\alpha +b_{0}t}{2}\right)  \label{Sec4-03}
\end{equation}%
From the definition of weak value, one can find that if the free evolution
of the spin is not taken into account, the resulting weak value is $%
i\cot\left(\alpha/2\right)$, rather than the above result.

It is well known that the weak value is linked to the final read of the
meter. To obtain the mean position, one should first expand $\left\langle
\psi_{{\rm fi}}\right\vert \exp\left(-i\xi\sigma_{z}t\right)\left\vert
\psi_{{\rm in}}\right\rangle $ until its first order in $b_{1}t$. Actually, this
first-order expansion is valid in our system since we only consider the
short-time behavior and the the external magnetic field gradient $B_{1}$ in
the $x$-direction is very small; thus $b_{1}t\ll1$ is satisfied. Then we write
$\left\langle \psi_{{\rm fi}}\right\vert \sigma_{z}\left\vert
\psi_{{\rm in}}\right\rangle $ in terms of the weak value $\left\langle
\sigma_{z}\right\rangle _{w}$. Finally, we regroup it to an exponential
function $\left\langle \psi_{{\rm fi}}\right.\left\vert \psi_{{\rm in}}\right\rangle
\left(1-i\xi_{t}t\left\langle \sigma_{z}\right\rangle
_{w}\right)\approx\left\langle \psi_{{\rm fi}}\right.\left\vert
\psi_{{\rm in}}\right\rangle \exp\left(-i\xi_{t}t\left\langle
\sigma_{z}\right\rangle _{w}\right)$, which yields
\begin{equation}
\left\langle x\right\rangle _{wv}=a^{2}b_{1}t\cot \left(\frac{\alpha +b_{0}t}{2}\right)
\label{Sec4-04}
\end{equation}%
as the observed mean position of the meter.

\begin{figure}[tbp]
\includegraphics[width=8cm]{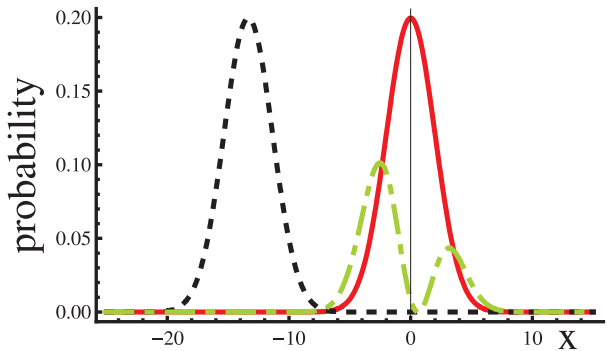}
\caption{(Color online) Transverse distribution of the probe field. The
red solid curve corresponds to the Gaussian profile in Eq.~(\ref{Sec3-08}).
The black dashed curve depicts the norm square of the wavefunction
in Eq.~(\ref{Sec4-06}) with a weak measurement right after the
light leaves the medium. The green dashed-dotted curve presents the normalized
distribution of Eq.~(\ref{Sec3-04}) at time $t_{f}=L/c$. Here, $x$ is
in units of millimeters. $b_{0}t+\alpha =0.08$. Other parameters are
the same as in Fig.~\protect\ref{fig3:1}.}
\label{fig4:1}
\end{figure}
It can be seen from Eq.~(\ref{Sec4-04}) that the final read of the meter is
proportional to the imaginary part of the weak value. To find the relation
between the result in Eq.~(\ref{Sec3-11}) and the weak value in Eq.~(\ref%
{Sec4-04}), we rewrite Eq.~(\ref{Sec3-03}) as
\begin{equation}
\left\vert E_{j}(t)\right\rangle =\exp[-i\left( cp_{z}+b_{0j}\right) t]\left(
1-ib_{1j}xt\right) \left\vert E_{0}(0)\right\rangle  \label{Sec4-05}
\end{equation}%
by retaining the linear term of the Taylor series expansion of $\exp \left(
-ib_{1j}xt\right) $. After some algebra, we find that the mean position in
Eq.~(\ref{Sec4-04}) is a linear approximation of Eq.~(\ref{Sec3-11}) with
respect to the coupling between the system and the meter. After the
postselection of the polarization degrees of freedom, the normalized
wavepacket of the probe field in the transverse direction becomes
\begin{equation}
\Psi _{f}^{N}\left( x,t\right) =\frac{1}{\sqrt{2\pi a^{2}}}\exp\left(-\frac{%
x^{2}-x\left\langle x\right\rangle _{wv}+\left\langle x\right\rangle
_{wv}^{2}}{4a^{2}}\right)\text{.}  \label{Sec4-06}
\end{equation}%
In Fig.~\ref{fig4:1}, we plot the transverse distribution of the incident
wavepacket in Eq.~(\ref{Sec3-08}) with a red solid curve, $|\Psi_{f}^{N}\left( x,t\right) |^{2}$
in Eq.~(\ref{Sec4-06}) at time $t_{f}=L/c$ (black dashed curve), and the
normalized norm square of Eq.~(\ref{Sec3-04}) right after the probe field
leaves the atomic medium (green dashed-dotted curve). Obviously, the weak
measurement significantly enhances the deflection of the probe field.

Different from the mean value of a quantum-mechanical measurement, which must
lie within the range of eigenvalues, weak values in Eq.~(\ref{Sec4-03})
produce results much larger than any of the eigenvalues of an observable,
particularly when one chooses the intial state $\left\vert \psi_{{\rm in}}
\right\rangle $ with $\alpha =-b_{0}t_{f}$ (where $t_{f}$ is the total
interaction time). By performing a weak measurement of the probe laser field
that has passed through an EIT atomic medium, we are able to significantly
magnify the transverse displacement of the probe field, which results in a
large group velocity $d\left\langle x\right\rangle _{wv}/dt$ in
the transverse direction. The deflection angle given in Ref.~\cite{Karpa06}
is defined as $\theta =c^{-1}d\left\langle x\right\rangle _{wv}/dt$. When
$b_{0}=0$, the deflection angle is totally decided by the original
polarization angle $\alpha $. And the magnitude of the deflection angle
could be arbitrarily large as $\alpha $ approaches zero. Comparing to the
angle of deflection $2\times 10^{-5}$ rad for the EIT condition~\cite{Karpa06}, the
weak measurement technique discussed here drastically amplifies the
displacement of the probe field in the transverse direction.

\section{\label{Sec:5}discussion}

We have theoretically studied a magneto-optically-controlled atomic ensemble
under the EIT condition to couple a property (the observable $\sigma _{z}$) of the
polarization (the system) with the spatial degree of freedom (the meter). In
the paraxial regime, the dynamics of the transverse distribution is governed
by an impulsive measurement interaction Hamiltonian $H_{{\rm int}}=\xi\sigma _{z}$,
which makes the displacements of the transverse spatial components
polarization-dependent. An enhanced displacement in the meter distribution is
achieved by an appropriate preselection and postselection of the polarization
state. However, the choice of the preselected state is dependent on the accumulated
phase during the free evolution of the system with the weak measurement taking
place in between.

This work is supported by NSFC No.~11074071, NFRPC~2012CB922103, PCSIRT No.~IRT0964,
Hunan Provincial Natural Science Foundation of China (11JJ7001 and 12JJ1002).
FN is partially supported by the ARO, JSPS-RFBR contract No.~12-02-92100,
Grant-in-Aid for Scientific Research (S), MEXT Kakenhi on Quantum Cybernetics,
and the JSPS via its FIRST program.


\begin{thebibliography}{99}
\bibitem{WVPRL60} Y. Aharonov, D.Z. Albert, and L. Vaidman, Phys. Rev. Lett.
\textbf{60}, 1351 (1988).

\bibitem{qparadox} Y. Aharonov, D. Rohrlich, Quantum Paradoxes-
Quantum Theory for the Perplexed (WILEY-VCH Verlag GmbH \& Co. KGaA).

\bibitem{VonNm} J.V Neumann, Mathematical Foundations of Quantum Mechanics
(Princeton University Press, Princeton 1955); German Version 1932.

\bibitem{WVpr12} A.G. Kofman, S. Ashhab, and F. Nori, Phys. Rep. \textbf{%
520}, 43 (2012).

\bibitem{phoPRL91} N.W.M. Ritchie, J.G. Story, and R.G. Hulet, Phys. Rev.
Lett. \textbf{66}, 1107 (1991).

\bibitem{phoPRL05} G.J. Pryde, J.L. O'Brien, A.G. White, T.C. Ralph, and
H.M. Wiseman, Phys. Rev. Lett. \textbf{94}, 220405 (2005).

\bibitem{SagPRL09} P.B. Dixon, D.J. Starling, A.N. Jordan, and J.C. Howell,
Phys. Rev. Lett. \textbf{102}, 173601 (2009).

\bibitem{GCGuoA06} Q. Wang, F.-W. Sun, Y.-S. Zhang, Jian-Li, Y.-F. Huang,
G.-C. Guo, Phys. Rev. A \textbf{73}, 023814 (2006).

\bibitem{WilRL08} N. S. Williams and A. N. Jordan, Phys. Rev. Lett. \textbf{%
100}, 026804 (2008).

\bibitem{RomPRL08} A. Romito, Y. Gefen, and Y.M. Blanter, Phys. Rev. Lett.
\textbf{100}, 056801 (2008).

\bibitem{lumPRL04} N. Brunner, V. Scarani, M. Wegm\"{u}ller, M. Legr\'{e}
and N. Gisin, Phys. Rev. Lett. \textbf{93}, 203902 (2004).

\bibitem{Bliokh06} K.Y. Bliokh and Y.P. Bliokh, Phys. Rev. Lett. 96, 073903 (2006).

\bibitem{OSHE08} O. Hosten and P. Kwiat, Science \textbf{319}, 787 (2008).

\bibitem{OSHE08F} F. Nori, Nature Photon. \textbf{2}, 716 (2008).

\bibitem{OSHE08n} K.Y. Bliokh, A. Niv, V. Kleiner, and E. Hasman, Nature Photon. \textbf{2}, 748 (2008).

\bibitem{Gorod12} Y. Gorodetski, K. Y. Bliokh, B. Stein, et al., Phys. Rev. Lett. \textbf{109}, 013901 (2012).

\bibitem{Karpa06} L. Karpa and M. Weitz, Nature Phys. 2, 332 (2006)

\bibitem{DLZhou07} D. L. Zhou, L. Zhou, R. Q. Wang, S. Yi, and C. P. Sun,
Phys. Rev. A 76, 055801 (2007).

\bibitem{ZhouA08} L. Zhou, J. Lu, D. L. Zhou, and C. P. Sun, Phys. Rev. A
\textbf{77}, 023816 (2008);

\bibitem{ZhangA09} H. R. Zhang, L. Zhou, and C. P. Sun, Phys. Rev. A \textbf{%
80}, 013812 (2009).

\bibitem{GuoA08} Y. Guo, L. Zhou, L. M. Kuang, and C. P. Sun, Phys. Rev. A
\textbf{78}, 013833 (2008);

\bibitem{Lukin} M. Fleischhauer and M. D. Lukin, Phys. Rev. Lett. \textbf{84}%
, 5094 (2000).

\bibitem{Lukin1} M. Fleischhauer and M. D. Lukin, Phys. Rev. A \textbf{65},
022314 (2002).

\bibitem{SunPRL01} C. P. Sun, Y. Li, and X. F. Liu, Phys. Rev. Lett. \textbf{%
91}, 147903 (2003).

\bibitem{Petrosyan} D. Petrosyan and Y. P. Malakyan, Phys. Rev. A \textbf{70}%
, 023822 (2004).

\bibitem{KarpaPRL} L. Karpa, F. Vewinger, and M. Weitz, Phys. Rev. Lett.
\textbf{101}, 170406 (2008).

\bibitem{Aharonov90} Y. Aharonov and L. Vaidman, Phys. Rev. A 41, 11 (1990).
\end{thebibliography}
\end{document}